\documentclass[aps,pra,twocolumn,showpacs,superscriptaddress]{revtex4}
\usepackage{graphicx}
\usepackage{amsmath}
\usepackage{amssymb}

\input{epsf}
\begin{document}

\title{Tuning the structural 
and dynamical properties 
of a dipolar Bose-Einstein condensate: Ripples and
instability islands}

\author{M. Asad-uz-Zaman}
\affiliation{Department of Physics and Astronomy,
Washington State University,
  Pullman, Washington 99164-2814, USA}
\author{D. Blume}
\affiliation{Department of Physics and Astronomy,
Washington State University,
  Pullman, Washington 99164-2814, USA}

\date{\today}

\begin{abstract}
It is now well established that the
stability of aligned dipolar Bose gases can be tuned by
varying the aspect ratio of the external harmonic confinement.
This paper extends this idea and demonstrates that 
a Gaussian barrier along the strong confinement direction
can be employed to tune both the structural properties 
 and the
dynamical stability of an oblate dipolar
Bose gas aligned along the strong confinement direction.
In particular,
our theoretical mean-field analysis predicts 
the existence 
of instability islands
immersed in otherwise stable regions of the phase diagram.
Dynamical studies indicate that these
instability islands, which can be probed experimentally with present-day
technology, are associated with the going soft of a Bogoliubov--de Gennes
excitation frequency with radial breathing mode character.
Furthermore, we find
dynamically stable 
ground state densities 
with 
ripple-like oscillations along the radial direction.
These structured ground states exist in the vicinity of a dynamical
radial roton-like instability.
\end{abstract}


\maketitle
\section{Introduction}
\label{introduction}
Dipolar Bose Einstein condensates (BECs)
such as magnetic Cr condensates are characterized by
angle-dependent long-range 
interactions~\cite{grie05,bara08,laha08a}. 
Usually, the magnetic dipole-dipole interactions
compete 
with 
the isotropic short-range $s$-wave contact interactions,
which dictate the behaviors of alkali atom BECs~\cite{dalf98} 
as well as of two-component
Fermi gases such as $^{40}$K and $^{6}$Li~\cite{gior08,kett08}.
In Cr, the
$s$-wave scattering
length $a_s$ can be tuned to zero through the 
application of an external field in the vicinity
of a Fano-Feshbach resonance, thus allowing for 
the realization 
of 
a pure dipolar
Bose gas~\cite{wern05,koch08,laha08}.
Although seemingly simple, dipolar
Bose gases in which the dipoles are aligned along
a particular laboratory axis
have been shown to exhibit a variety of intriguing and unique
features such as so-called roton instabilities~\cite{sant03,dell03} and 
structured density profiles~\cite{rone07,dutt07,wils08}. 
Importantly,
although the strength of the magnetic dipole-dipole interactions is fixed
by the electronic structure of the atoms, 
it can be 
tuned effectively by varying the
aspect ratio of the 
external confining potential~\cite{sant00,metz09}. 
This effective tunability of the dipolar interactions 
underlies a series of studies of dipolar
gases and can be readily
understood intuitively.
Assuming the dipoles are aligned along the $z$-axis,
an effectively one-dimensional confinement that predominantly
allows for motion along the $z$-axis leads 
to an attraction between the dipoles and 
thus collapse. An effectively two-dimensional confinement
that predominantly allows for motion
in the $xy$-plane, in contrast, leads to a repulsion
between the dipoles and thus stabilization.

This paper considers a pure dipolar Bose gas, aligned along the
$z$-axis, in an oblate trapping geometry with cylindrical
symmetry characterized by the aspect ratio $\lambda$. 
In addition, a repulsive 
Gaussian barrier with fixed width $b$
and variable height $A$ is added
along the $z$ direction
(centered at $z=0$). 
When the chemical potential $\mu$ is much larger than the barrier
height $A$, the Gaussian potential serves as a small perturbation.
For $\mu/A \ll 1$, in contrast, the fairly narrow Gaussian barrier 
significantly modifies the system behavior and 
approximately splits
the dipolar condensate
into two ``sheets'' or ``pancakes''.
We determine the phase and stability diagram of the
dipolar gas as a function of the mean-field strength $D$
and the barrier height $A$ for various 
aspect ratios $\lambda$. 
Our key results are as follows:
(i) We find an instability island immersed in otherwise stable regions of the
$D$ versus $A$
phase diagram. Our analysis of the Bogoliubov--de Gennes excitation
frequencies and corresponding eigenmodes indicates that these
unstable islands arise due to a breathing mode-like instability.
(ii) We find dynamically stable density profiles of Gaussian shape with
superimposed ripple-like structure. For certain parameter 
combinations near the instability line,
these ripple-like structures
exist for vanishing as well as finite Gaussian barrier.
As the mean-field strength increases, one of the Bogoliubov--de 
Gennes frequencies with radial roton type character becomes 
soft, inducing a dynamical
instablility.

\begin{figure}
\vspace*{.2cm}
\includegraphics[angle=0,width=70mm]{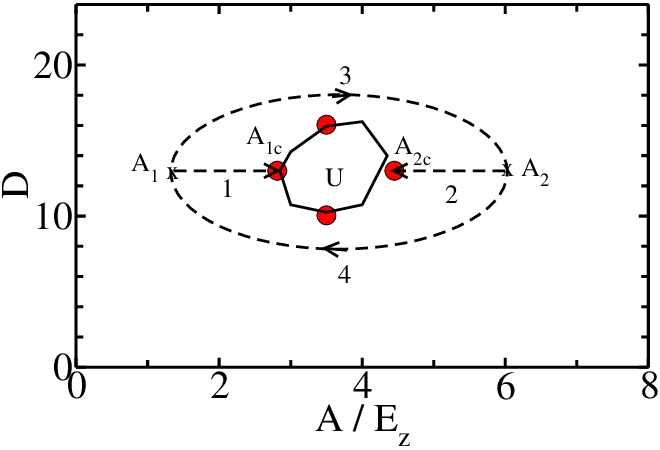}
\caption{
(Color online)
Probing the instability island.
A solid line separates
the mechanically stable region and the mechanically unstable 
island (labeled ``U'') in the mean-field strength $D$
versus barrier height $A$ phase
diagram for $\lambda=8$ and $b=0.2a_z$.
Filled circles indicate $(D,A)$ values at which the $k=0$ mode
becomes dynamically unstable  (see text for details).
The instability island can be probed experimentally by
following trajectories 1 and 2 that end at the edges of the
dynamically unstable island (labeled $A_{1c}$ and $A_{2c}$), 
and by following trajectories 3 and 4 that
encircle the instability island.
}\label{fig_trajectory}
\end{figure}

Experimentally, the 
instability island 
predicted by our calculations
can be probed straightforwardly by following two different
types of trajectories
in the $D$ versus $A$ phase diagram:
(i) The first experiment probes the instability by 
adiabatically increasing the barrier height 
from $A_{1}$ to $A_{1c}$
or by adiabatically decreasing the barrier height 
from $A_{2}$ to $A_{2c}$
for fixed aspect ratio
$\lambda$, fixed mean-field strength
$D$ and fixed barrier width $b$ (see trajectories 1 and 2 in
Fig.~\ref{fig_trajectory}). The experimental signature of the collapse
would be 
the onset of significant atom loss at $A_{1c}$ and $A_{2c}$, 
respectively. 
(ii) The second experiment by-passes 
the instability island by 
moving from $A_{1}$ to $A_{2}$ and back to $A_1$ by simultaneously
varying $A$ and $D$  
while keeping $\lambda$ and $b$
fixed (trajectories 3 and 4 in Fig.~\ref{fig_trajectory}
trace out one of many possible paths).
In the case of Cr, varying the mean-field strength $D$, 
which is determined by 
the number of particles $N$ and the square of
the dipole moment strength $d$, implies varying $N$.
Thus, trajectories 3 and 4
can be mapped out by preparing dipolar BECs with varying 
$N$ and observing long condensate lifetimes
for all $(D,A)$ parameter combinations along the
chosen trajectories.

The remainder of this paper is organized as follows.
Section~\ref{sec_theory} introduces the system under study and the
mean-field formalism.
Section~\ref{sec_results} presents our results
and
Sec.~\ref{sec_conclusion} 
concludes.

\section{Mean-field description of dipolar BECs}
\label{sec_theory}
Our description of the aligned dipolar Bose gas 
with vanishing $s$-wave interactions
is based on the 
time-dependent 
Gross-Pitaevskii equation~\cite{yi00,gora00,yi01}
\begin{eqnarray}
\label{eq_gp}
i \hbar \frac{\partial \Psi(\vec{r},t)}{\partial t} =
\bigg(
- \frac{\hbar^2}{2m}\nabla^2 + V_{\mathrm{ext}}(\vec{r}) + \nonumber  \\
d^2 (N-1)  \int \frac{1 - 3 \cos^2 \vartheta}{|\vec{r}-\vec{r}'|^3}
 |\Psi(\vec{r}',t)|^2 d^3 \vec{r}' \bigg)
\Psi(\vec{r},t),
\end{eqnarray}
where $m$ denotes the mass of the dipoles,
$d$ the strength of the
dipole moment, 
and $\vartheta$ the angle between the relative distance vector
$\vec{r}-\vec{r}'$ and the $z$-axis. 
Throughout, we employ cylindrical coordinates $\vec{r}=(\rho,\varphi,z)$.
The external
confining potential
$V_{\mathrm{ext}}(\rho,z)$,
\begin{eqnarray}
\label{eq_vho}
V_{\mathrm{ext}}(\rho,z) = \frac{1}{2} m (\omega_{\rho}^2 \rho^2 + \omega_z^2 z^2)
+
A \exp \left(- \frac{z^2}{2 b^2} \right),
\end{eqnarray}
is characterized by the transverse and longitudinal angular trapping
frequencies $\omega_{\rho}$ and $\omega_z$,
respectively, 
and the height $A$ and width $b$ of the repulsive Gaussian
barrier.
Throughout, we fix $b$, i.e., we set $b=0.2a_z$,
where $a_z = \sqrt{{\hbar}/(m \omega_z)}$.
The angular trapping frequencies define
the aspect ratio $\lambda$,
\begin{eqnarray}
\label{eq_lambda}
\lambda = \frac{\omega_z}{\omega_{\rho}}.
\end{eqnarray}
Throughout, we consider oblate systems with $\lambda=6-9$.
If the Gaussian barrier is sufficiently large, the
system consists---to lowest order---of two 
staggered pancakes; in this case, we define an
effective aspect ratio $\lambda_{\mathrm{eff}}$, which is obtained by analyzing
the energy spectrum for large $A$ and vanishing interactions, i.e., for
$D=0$
(see below).
To quantify the strength
of the mean-field term in Eq.~(\ref{eq_gp}), we define 
the dimensionless quantity $D$,
\begin{eqnarray}
\label{eq_dmf}
D= \frac{d^2(N-1)}{E_{\rho} a_{\rho}^3},
\end{eqnarray}
where $E_{\rho}=\hbar \omega_{\rho}$
and $a_{\rho} = \sqrt{{\hbar}/(m \omega_{\rho})}$.
Note that the quantity $D$ defined in Eq.~(\ref{eq_dmf}) differs
from the definition used in our previous work~\cite{asad09}
but agrees with that used in Ref.~\cite{rone07}.

We consider stationary and dynamical properties of the 
dipolar Bose gas.
Our stationary calculations determine the energetically lowest-lying
gas-like state of the dipolar system. We
write $\Psi(\vec{r},t)=\exp(-i \mu t/\hbar) \psi_0(\rho,\varphi,z)$,
where 
$\psi_0$ denotes
the stationary ground state solution and $\mu$ the 
corresponding chemical potential.
Since the system is cylindrically symmetric,
$\psi_0$ is a function of $\rho$ and $z$ only and not of $\varphi$.
We solve the resulting two-dimensional Schr\"odinger equation
by propagating a given initial state in imaginary time. In selected cases,
we additionally employ a basis set expansion type approach. 
We have checked that the results
of these two distinctly 
different approaches, whose implementation follows that
outlined 
in Ref.~\cite{rone07,modu03,asad09}, agree within our numerical 
uncertainties.

In addition to the ground state properties, we determine the 
Bogoliubov--de Gennes eigenfrequencies and eigenmodes, which provide
insights into the dynamical stability of the system~\cite{dalf98,dalf97}.
To this end, we solve the Bogoliubov--de Gennes eigenequations, 
which are derived
within linear response theory assuming that the ground state $\psi_0$
is perturbed by
$\delta \psi(\vec{r},t)$,
\begin{eqnarray}
\delta \psi(\vec{r},t) = 
u(\vec{r}) \exp(-i \omega t) +
v^*(\vec{r}) \exp(i \omega t),
\end{eqnarray}
iteratively using the Arnoldi method~\cite{rone06a}.
Here, $u(\vec{r})$ and $v(\vec{r})$ are the so-called Bogoliubov--de
Gennes functions and $\omega$ is
the Bogoliubov--de Gennes eigenfrequency.
Following the approach introduced in Ref.~\cite{rone06a},
we solve the Bogoliubov--de Gennes eigenequations
for $\omega^2$ and for the so-called Bogoliubov--de Gennes 
eigenmode $f(\vec{r})$, where $f(\vec{r})=u(\vec{r})+v(\vec{r})$.
Because of the cylindrical symmetry of
the system, the $\varphi$ dependence can be separated off,
i.e., $f(\vec{r})= \bar{f}(\rho,z) h(\varphi)$ with 
$h(\varphi)=\exp(ik\varphi)/\sqrt{2 \pi}$.
This implies that 
the Bogoliubov--de Gennes 
excitation frequencies $\omega$ 
can,
for a given
azimuthal quantum number $k$, 
be obtained by solving
a two-dimensional eigenvalue problem.
Typically, we consider excitations with up to $k=4$.
For the discussions
in Sec.~\ref{sec_results},
it is important to recall that a negative $\omega^2$,
and thus a purely imaginary $\omega$, signals a dynamically 
unstable ground state $\psi_0(\vec{r})$.
Correspondingly, an analysis of the Bogoliubov--de Gennes eigenmodes
$f(\vec{r})$ near the dynamical instability point provides
insights into the decay mechanism;
in particular, the 
correction to the
time-dependent density is, to lowest order in the perturbation
$\delta \psi$, given by 
$2\cos(\omega t) \psi_0(\vec{r}) f(\vec{r})$.

\begin{figure}
\vspace*{.2cm}
\includegraphics[angle=0,width=70mm]{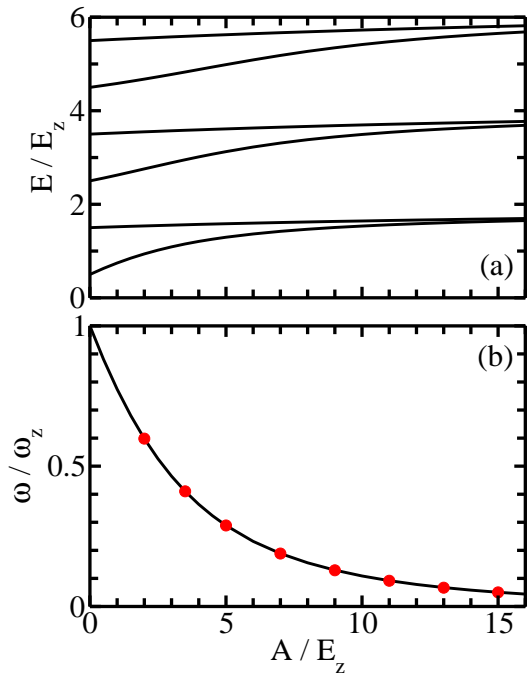}
\caption{
(Color online)
Energetics of the one-dimensional double-well potential.
Solid lines
show (a) the eigenenergies and (b) the lowest eigenfrequency
of the one-dimensional double
well potential as a function of the barrier height $A$.
The spectrum consists of nearly equally spaced energy 
levels for small $A$
and nearly equally spaced
energy level pairs
for large $A$.
For comparison, symbols in panel~(b) show 
the eigenfrequencies obtained by solving the Bogoliubov--de Gennes
equations for $D=0$;
excellent agreement between the two sets of frequencies is found.
}\label{fig_tunneling}
\end{figure}

To elucidate the energy and length scales of $V_{\mathrm{ext}}(\rho,z)$,
we consider the non-interacting system (i.e., we set $D=0$).
In this case, the linear Schr\"odinger equation separates
and the solutions are readily obtained.
In the $(\rho,\varphi)$-plane, we have a two-dimensional
harmonic oscillator with eigenenergies
$(2n_{\rho}+|k|+1)E_{\rho}$, where $n_{\rho}$ and
$k$ denote the principal and azimuthal quantum numbers, respectively 
($n_{\rho}=0,1,\cdots$ and $k=0,\pm 1,\cdots$).
In the $z$-direction, the energy spectrum 
depends on two parameters, the axial frequency $\omega_z$
and the barrier height $A$ (recall, the barrier width $b$ is kept
constant throughout this paper).
Figure~\ref{fig_tunneling}
shows the lowest six eigenenergies, obtained by
diagonalizing  the Hamiltonian matrix expressed in terms of harmonic
oscillator functions, as a 
function of $A$ for $b=0.2a_z$. 
For $A=0$, the eigenenergies follow the harmonic
oscillator pattern $(n_z+1/2)E_z$, where $n_z=0,1,\cdots$ and
$E_z=\hbar \omega_z$.
For large $A$, in contrast, the eigenenergies appear in
roughly equally spaced pairs [Fig.~\ref{fig_tunneling}(a)
shows three of these pairs].
We use the energy spacing between the lowest two pairs to define 
an effective angular frequency $\omega_{z,{\mathrm{eff}}}$ 
that characterizes the
left and the right well of the double well potential.
Extrapolating to the large barrier height limit, we find
$\omega_{z,{\mathrm{eff}}}\approx
2 \omega_z$. 
Correspondingly, we define an effective 
harmonic oscillator length $a_{z,{\mathrm{eff}}}$ through
$a_{z,{\mathrm{eff}}}=\sqrt{\hbar/(m \omega_{z,{\mathrm{eff}}})}$,
leading to $a_{z,{\mathrm{eff}}}\approx a_z/\sqrt{2}$.
While this analysis is approximate---our definition of 
$a_{z,{\mathrm{eff}}}$, e.g.,
assumes a harmonic potential around the minimum
of the left well and of the right well---, it provides insights into
how the energy and length scales 
associated with the $z$-degree of freedom
vary with increasing barrier height.

Lastly, we point out that the Gaussian barrier leads to an
energetically low-lying tunneling splitting mode for comparatively
large $A$ and $D=0$. 
For $A=16E_z$, e.g., we find that the density at $z=0$ is about 25 times
smaller than the peak density, suggesting that the left 
well and the right well are, to a good approximation,
decoupled.
Correspondingly, as shown in Fig.~\ref{fig_tunneling}(b), the lowest 
frequency decreases
from $1\omega_z$ for $A=0$ to about $0.044 \omega_z$
for $A=16E_z$.
This implies that
the tunneling splitting mode for $D=0$ is, in the 
large $A$ limit and for $\lambda=6-9$, energetically lower-lying than the 
lowest mode along the
$\rho$-direction which has a frequency of
$\omega_{\rho}$.
Our discussion of the excitation spectrum for vanishing $D$ presented here
serves as a guide to understanding the 
Bogoliubov--de Gennes excitation spectra for finite $D$ 
(see Sec.~\ref{sec_results}).

\section{Results}
\label{sec_results}
This section presents the energetics,
the density profiles, and selected Bogoliubov--de Gennes excitation
spectra and eigenmodes for 
trapped dipolar Bose gases.
In particular, we
detail
calculations
that predict an instability island 
(see Fig.~\ref{fig_trajectory})
and density profiles with ripple-like oscillations.

\begin{figure}
\vspace*{.2cm}
\includegraphics[angle=0,width=70mm]{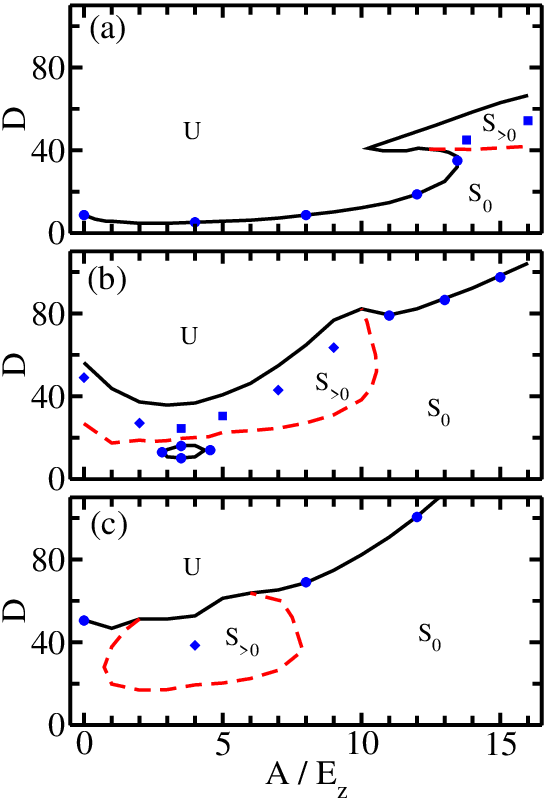}
\caption{
(Color online)
$D$ versus $A$
phase diagram for 
fixed $b$, $b=0.2a_z$, and three different
$\lambda$ values: (a) $\lambda=6$, (b) $\lambda=8$, and 
(c) $\lambda = 9$.
Solid lines
separate mechanically stable regions (labeled ``S$_0$'' and ``S$_{>0}$'') from
mechanically unstable regions (labeled ``U'').
The instability island centered around $A\approx 3.5E_z$
and $D\approx13$ for $\lambda=8$ 
[see panel~(b)] is shown on an enlarged scale 
in Fig.~\protect\ref{fig_trajectory}.
The dynamically stable regions
are separated from the dynamically
unstable regions by symbols: Filled circles,
squares and diamonds indicate that the
dynamical instability is triggered by a $k=0$, 2 and $3$ mode, respectively.
Dashed lines separate stationary ground state densities with maximum at
$\rho=0$ (labeled ``S$_0$'') from those with maximum at $\rho>0$ 
(labeled ``S$_{>0}$'').
In certain regions of the phase diagram, the ground state densities
possess ripple-like oscillations (the corresponding
phase boundaries are not indicated here; 
see text and Fig.~\protect\ref{fig_ripple2}).
}\label{fig_phasediagram}
\end{figure}

To understand how the instability island for $\lambda=8$ emerges,
Figs.~\ref{fig_phasediagram}(a)-(c)
show the $D$ versus $A$ phase diagram for fixed $b$ and three different
$\lambda$ values, i.e., for $\lambda=6$, $8$ and $9$.
Compared
to Fig.~\ref{fig_trajectory},
the phase diagram for $\lambda=8$ [see Fig.~\ref{fig_phasediagram}(b)]
shows an extended $(D,A)$ parameter regime.
As in Fig.~\ref{fig_trajectory}, 
mechanically stable regions of the phase diagram are separated
by solid lines from mechanically unstable 
regions of the phase diagram. 
Dynamically stable regions are separated from 
dynamically unstable regions by
symbols; filled circles, squares and 
diamonds indicate that the dynamical instability
is triggered by a $k=0$, $2$ and $3$ mode, respectively.
The mechanical and dynamical 
instability points that separate the ``upper'' and 
``lower'' 
regions of the phase diagram differ by up to
about 30~\%. Along these instability
lines, the decay is triggered by modes with vanishing or non-vanishing 
azimuthal quantum number $k$
and
has previously been discussed 
for
$A=0$ in
Ref.~\cite{rone07} and for $A=12E_z$ in Ref.~\cite{asad09}.
In the vicinity
of the instability island,
which is centered at $A \approx 3.5E_z$
and $D \approx 13$
[see elliptically-shaped solid lines in 
Fig.~\ref{fig_phasediagram}(b) and Fig.~\ref{fig_trajectory}], the 
$k=0$ Bogoliubov--de Gennes 
mode 
becomes soft first; in this region,
the dynamically unstable region is only
slightly larger than the mechanically unstable region. 

Inspection of Figs.~\ref{fig_phasediagram}(a)-(c) shows that
the
instability island
for $\lambda=8$ emerges from the break-up
of the unstable region of the phase diagram for $\lambda=6$
into two pieces
[see Fig.~\ref{fig_phasediagram}(a)].
For $\lambda=7$ (not shown), 
we find a somewhat larger mechanically unstable island
than for $\lambda=8$,
centered around roughly the
same $(D,A)$ values as
for $\lambda=8$.
Thus,
as $\lambda$ increases
the instability island shrinks, and it is absent
for $\lambda=9$.
Figure~\ref{fig_phasediagram}(b) shows that the instability island
is surrounded by a dynamically stable region, 
suggesting 
that the
experimental sequence discussed in the introduction in the
context of Fig.~\ref{fig_trajectory} is feasible
with present-day technology. Specifically, 
for a Cr condensate with
vanishing $s$-wave scattering length,
confined by a trap with 
$\omega_{\rho}=2 \pi \times 100$~Hz and $\omega_z=2 \pi \times 800$~Hz, 
point $A_1$  in Fig.~\ref{fig_trajectory} corresponds
to $N \approx 7200$ atoms, while the 
dynamical instability points for $A=3.5E_z$
in Fig.~\ref{fig_phasediagram}(b)
correspond to (from bottom to top)
$N \approx 5900$, 9300 and 14200.

In determining the $D$ versus $A$ phase diagrams shown in 
Figs.~\ref{fig_phasediagram}(a)-(c),
our ground state calculations employed a 
grid with widths
$4$ and $1E_z$ in $D$ and $A$, respectively. 
In a second set of calculations, a finer grid in $D$ was
employed to determine the mechanical instability point more 
accurately and to analyze the dynamical stability.
We cannot rule 
out the existence of additional instability islands with smaller
widths 
than our numerical resolution.

\begin{figure}
\vspace*{.2cm}
\includegraphics[angle=0,width=70mm]{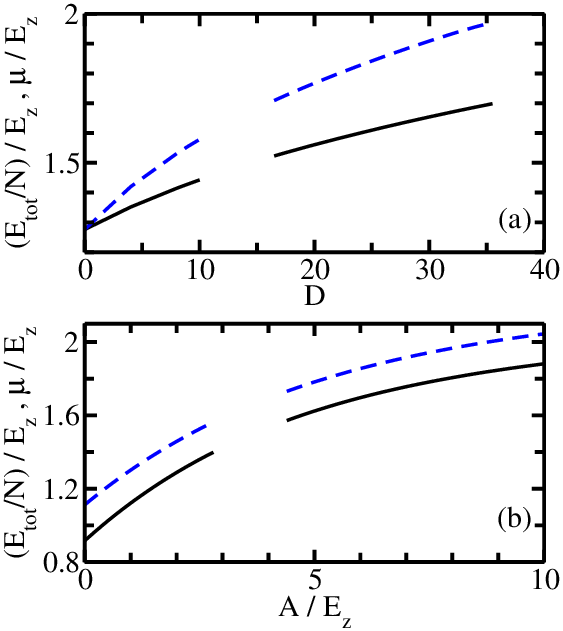}
\caption{
(Color online)
Energetics across the instability island.
Solid and dashed lines show
the total energy per particle 
$E_{\mathrm{tot}}/N$ and the chemical potential $\mu$, respectively,
for $\lambda=8$ and $b=0.2a_z$.
In panel~(a), the energetics are shown as a function of 
$D$ for $A=3.5E_z$; in panel~(b), the energetics are shown as
a function of $A$ for $D=13$.
}\label{fig_energetics}
\end{figure}

Figure~\ref{fig_energetics}
shows the energy per particle $E_{\mathrm{tot}}/N$ 
(solid line) and the chemical potential
$\mu$ (dashed line) for $\lambda=8$ and $b=0.2a_z$:
Fig.~\ref{fig_energetics}(a) shows the energetics 
for $A=3.5E_z$ (corresponding to $28E_{\rho}$) as a function of $D$
[i.e., along a vertical cut in Fig.~\ref{fig_phasediagram}(b)], 
while Fig.~\ref{fig_energetics}(b) shows the
energetics for $D=13$ as a function of $A$
[i.e., along a horizontal cut in Fig.~\ref{fig_phasediagram}(b)].
For fixed $D$
[see Fig.~\ref{fig_energetics}(b)], 
the energy and chemical potential increase monotonically
till the mechanical instability 
point is reached at $A\approx 2.85E_z$. At this point,
the chemical potential $\mu$ is about two times smaller
than the barrier height.
Beyond the mechanically unstable region, i.e., for $A \gtrsim 4.4E_z$,
both $E_{\mathrm{tot}}/N$ and $\mu$ continue to increase monotonically.
The energy per particle and the chemical potential
behave similarly as a function of $D$ 
for fixed $A$ [see Fig.~\ref{fig_energetics}(a)].
Interestingly, the presence of the instability island manifests itself
in the form of a ``gap'' in the energy and the chemical potential
but does not notably change the slope of either of these quantities
(i.e., $E_{\mathrm{tot}}/N$ and $\mu$ appear to be changing smoothly across
the instability island). We find a similar behavior for a subset of the
Bogoliubov--de Gennes eigenmodes (see below).

\begin{figure}
\vspace*{.2cm}
\includegraphics[angle=0,width=70mm]{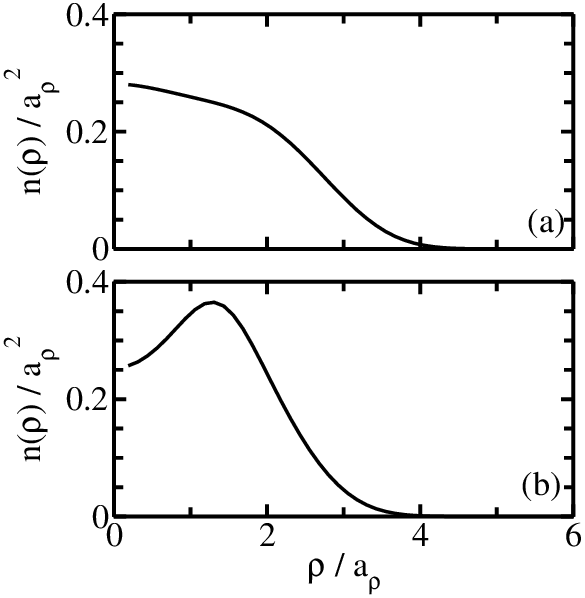}
\caption{
Examplary integrated density 
profiles ${n}(\rho)$
as a function of $\rho$ for $\lambda=9$, $b=0.2a_z$ and 
two different 
$(D,A)$ combinations:
(a) $(D,A)=(40,0)$ (S$_0$ type density) and
(b) $(D,A)=(36,4E_z)$ (S$_{>0}$ type density).
Both density profiles correspond to dynamically stable dipolar Bose gases
near the instability line [see Fig.~\protect\ref{fig_phasediagram}(c)].
}\label{fig_ripple}
\end{figure}

In addition to the energetics, we analyze 
the structural properties of dipolar
Bose gases in a double-well potential.
Dashed lines in Figs.~\ref{fig_phasediagram}(a)-(c)
separate the parameter region where 
the ground state density $|\psi_0|^2$ has its maximum at 
$\rho=0$ 
(labeled S$_0$) from that where $|\psi_0|^2$ has its maximum at $\rho>0$
(labeled S$_{>0}$). The transition from the S$_0$ type
to the S$_{>0}$ type density
is smooth and the dashed lines are based on an analysis of 
the integrated density ${n}(\rho)$,
${n}(\rho)=2 \pi \int |\psi_0(\rho,z)|^2 dz$.
We denote the $\rho$ value at which
${n}(\rho)$ is maximal by $\rho_{\mathrm{max}}$ and define
the transition from S$_0$ to S$_{>0}$ type densities, i.e., the dashed lines
in Fig.~\ref{fig_phasediagram},
by the condition ${n}(0)/{n}(\rho_{\mathrm{max}})=0.98$.
Figures~\ref{fig_ripple}(a) and (b) show examplary integrated
densities ${n}(\rho)$ of types S$_0$ and S$_{>0}$,
respectively.
The latter density profile has, for a system
with vanishing barrier, been previously termed 
``red blood cell''~\cite{rone07};
this name is motivated by the fact that the system's isodensity plot
resembles the shape of a red blood cell.
For all parameter combinations investigated,
the dynamical instability is triggered by a $k=0$ mode
if the ground state density is of S$_0$ type and by a finite
$k$ mode if the ground state density is of S$_{>0}$ type.
In particular,
Fig.~\ref{fig_phasediagram}(b) shows that the density profiles in
the vicinity of the instability island have a simple Gaussian shape
and that the dynamical instability 
in this regime is associated with a $k=0$ mode.

\begin{figure}
\vspace*{.2cm}
\includegraphics[angle=0,width=70mm]{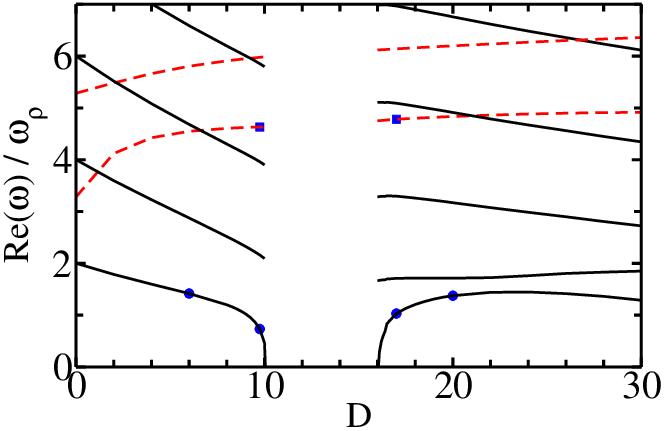}
\caption{
(Color online)
Bogoliubov--de Gennes
 eigenspectrum as a function
of $D$ for $k=0$, $A=3.5E_z$, $\lambda=8$ and $b=0.2a_z$.
Solid lines show frequencies for which the corresponding eigenmodes
$\bar{f}$
have nodal lines that are
parameterized by $\rho \approx \mbox{constant}$.
Dashed lines show frequencies for which the corresponding eigenmodes
have, in addition to other nodal lines, a nodal line given by $z=0$.
Filled circles and squares 
mark those $D$ values for
which Fig.~\ref{fig_bdgeigenmode} 
and Fig.~\ref{fig_bdgeigenmode2}, respectively,
show eigenmodes.
}\label{fig_bdgfrequency2}
\end{figure}

\begin{figure}
\vspace*{.2cm}
\includegraphics[angle=0,width=80mm]{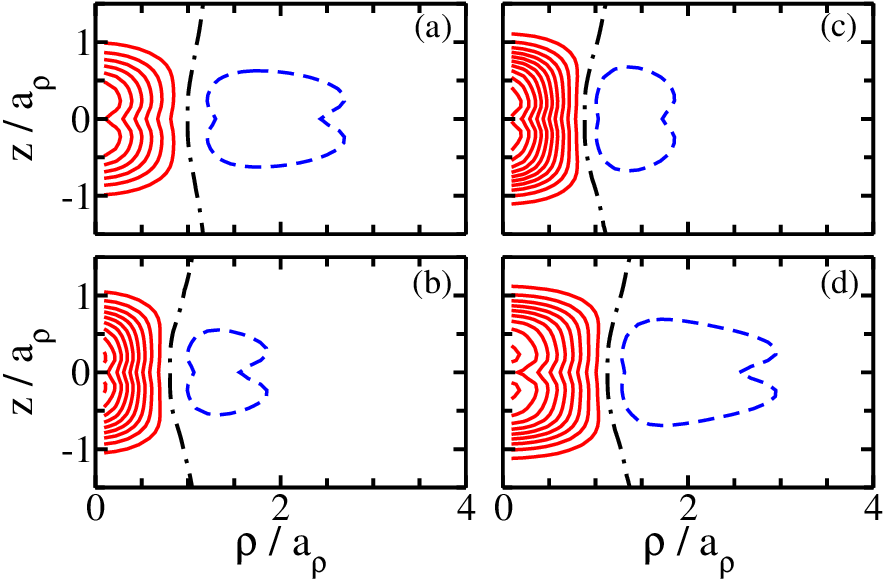}
\caption{
(Color online)
Examplary Bogoliubov--de Gennes eigenmodes 
with breathing mode character
for $k=0$, $A=3.5E_z$, $\lambda=8$, $b=0.2a_z$
and 
(a) $D=6$,
(b) $D=9.75$,
(c) $D=17$, and
(d) $D=20$.
The corresponding Bogoliubov--de Gennes eigenfrequencies are
shown by circles in
Fig.~\protect\ref{fig_bdgfrequency2} and 
in Fig.~\protect\ref{fig_bdgfrequency}.
The contours are chosen equidistant, with solid and dashed lines
corresponding to positive and negative values 
of $\bar{f}$. The dash-dotted lines indicate the nodal lines of $\bar{f}$.
}\label{fig_bdgeigenmode}
\end{figure}

\begin{figure}
\vspace*{.2cm}
\includegraphics[angle=0,width=80mm]{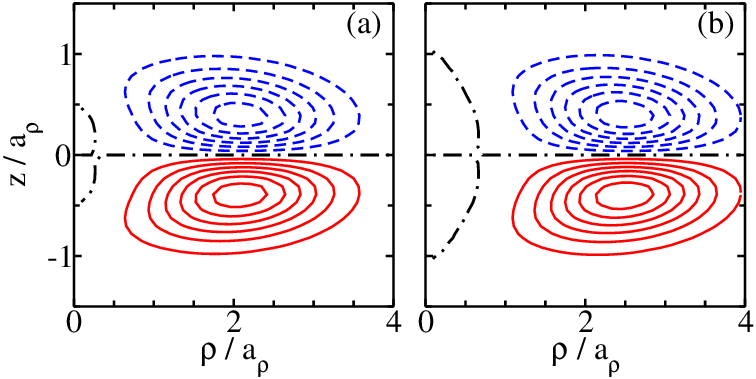}
\caption{
(Color online)
Examplary Bogoliubov--de Gennes eigenmodes $\bar{f}(\rho,z)$
with tunneling splitting character
for $k=0$, $A=3.5E_z$, $\lambda=8$, $b=0.2a_z$
and 
(a) $D=9.75$ and
(b) $D=17$.
The corresponding Bogoliubov--de Gennes eigenfrequencies are
shown by squares in
Fig.~\protect\ref{fig_bdgfrequency2}.
The contours are chosen equidistant, with solid and dashed lines
corresponding to positive and negative values 
of $\bar{f}$. The dash-dotted lines indicate the nodal lines of $\bar{f}$.
}\label{fig_bdgeigenmode2}
\end{figure}

To shed light on the dynamics in the vicinity 
of the instability island,
Fig.~\ref{fig_bdgfrequency2} 
shows the Bogoliubov--de Gennes excitation spectrum 
as a function of $D$ for
$k=0$, $A=3.5E_z$, $\lambda=8$ and $b=0.2a_z$.
The lowest excitation frequency becomes purely imaginary at $D\approx 10$,
signaling the dynamical instability.  
The frequency ``recovers'' at somewhat larger $D$ values
($D \approx 16$).
We find that the real parts of the
eigenfrequencies with non-vanishing azimuthal
quantum number $k$ remain finite in the vicinity of
the instability island,
implying that the dynamics
in this parameter space of the phase diagram
is dominated by $k=0$ modes.
Our analysis of the Bogoliubov--de Gennes eigenmodes
(see Figs.~\ref{fig_bdgeigenmode} and \ref{fig_bdgeigenmode2}
as well as the discussion below)
shows that
the eigenmodes $\bar{f}$ corresponding to the
Bogoliubov--de Gennes
excitation frequencies 
shown by solid lines in Fig.~\ref{fig_bdgfrequency2}
have nodal lines 
that are to a good approximation independent of
$z$
while those  corresponding to the
Bogoliubov--de Gennes
excitation frequencies 
shown by dashed lines have, among other nodal lines, a nodal line
given by $z=0$.
The
eigenfrequencies shown by solid lines
are affected by the instability (i.e., the slope
of these frequencies
changes near the dynamical instability),
while those shown by dashed lines
appear to change smoothly across
the instability.
This suggests that these two
classes of eigenfrequencies are approximately decoupled.
We find that
the finite $k$ frequencies also change as though
they are uneffected by the instability island.
For larger $D$ values ($D \gtrsim 24.5$),
however, modes with finite azimuthal quantum number become relevant
in determining the system's dynamical stability
for a fairly wide range of $A$ values.
In fact, as indicated in Fig.~\ref{fig_phasediagram}(b),
the instability for comparatively
large $D$ values
and $A=0$ to $A \approx 10E_z$ is triggered by $k=2$ and 3 modes 
(i.e., for these $A$ values the energetically
lowest-lying $k=2$ or 3 frequency becomes
purely imaginary
before the lowest $k=0$ frequency does).

\begin{figure}
\vspace*{.2cm}
\includegraphics[angle=0,width=70mm]{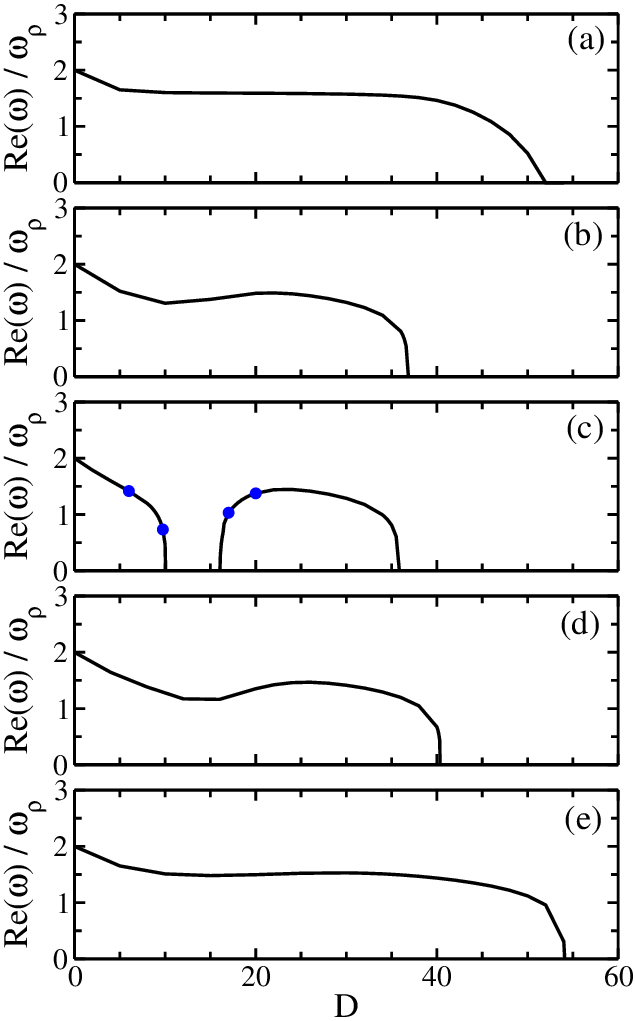}
\caption{
(Color online)
Real part of the energetically lowest-lying
Bogoliubov--de Gennes
 eigenfrequency $\omega$ as a function
of $D$ for $k=0$, $\lambda=8$, $b=0.2a_z$ and five different $A$ values:
(a) $A=0$,
(b) $A=2E_z$,
(c) $A=3.5E_z$,
(d) $A=5E_z$, and
(e) $A=7E_z$.
Filled circles in panel~(c) 
mark those $D$ values for
which Fig.~\ref{fig_bdgeigenmode} shows eigenmodes.
}\label{fig_bdgfrequency}
\end{figure}

Figure~\ref{fig_bdgfrequency} shows the real part of the energetically
lowest-lying $k=0$ 
Bogoliubov--de Gennes eigenfrequency
as a function of $D$ for $\lambda=8$, $b=0.2a_z$ 
and various $A$ values.
The frequency shown in Fig.~\ref{fig_bdgfrequency}(c)
for $A=3.5E_z$
has just been discussed in the context of 
Fig.~\ref{fig_bdgfrequency2}.
Inspection 
of Fig.~\ref{fig_bdgfrequency}
shows that the vanishing of the real part of the 
energetically lowest-lying breathing
mode type frequency for $A=3.5E_z$
is accompanied by a ``dip'' in the 
corresponding frequencies 
for smaller and larger $A$ [see Figs.~\ref{fig_bdgfrequency}(b) and (d)]
around $D \approx 10$ 
and 15, respectively.
For these $A$ values, however, the mode ``recovers'' as $D$ increases
and the system is mechanically and dynamically stable up to 
comparatively large $D$ values.

We now discuss the Bogoliubov--de Gennes eigenmodes 
shown in Fig.~\ref{fig_bdgeigenmode} and in Fig.~\ref{fig_bdgeigenmode2}
in more 
detail.
The $k=0$ eigenmodes shown in Fig.~\ref{fig_bdgeigenmode2}
show one nodal line that is given by $z=0$ and a second
nodal line that depends on $\rho$ and $z$.
The former nodal line reflects the symmetry of the confining
geometry and corresponds
to oscillations between the left well and the right well,
while the latter is of secondary
importance since the amplitude of $\bar{f}$ is small along this nodal line.
We classify the eigenfrequencies associated with eigenmodes of the type
shown in Fig.~\ref{fig_bdgeigenmode2} as having tunneling 
splitting mode character.
The $k=0$ eigenmodes shown in Fig.~\ref{fig_bdgeigenmode},
in contrast, are characterized by a single nodal line which
depends to first order only on $\rho$ and not on $z$.
We classify the eigenfrequencies associated with eigenmodes
of the type shown in Fig.~\ref{fig_bdgeigenmode} as having
breathing mode character.
An analysis of the eigenmodes
corresponding to the
energetically lowest-lying $k=0$ frequency
for other $D$ values
but the same $A$, $\lambda$ and $b$ shows that
the nodal line is located at $\rho=a_{\rho}$ for
$D=0$ and bends slightly as $D$ increases while remaining
located around $\rho \approx a_{\rho}$
[see Fig.~\ref{fig_bdgeigenmode}(a)-(b)]. 
Past the instability, the nodal line of $\bar{f}$
is again located at $\rho \approx a_{\rho}$
[see Fig.~\ref{fig_bdgeigenmode}(c)]
and  
moves to larger $\rho$ values
with increasing $D$
[see Fig.~\ref{fig_bdgeigenmode}(d)].
Eventually, a new nodal line moves in from 
$\rho=0$.
Near the dynamical instability at $D \approx 36$, the eigenmode
is characterized by two 
approximately equally spaced nodal lines 
with approximately constant $\rho$ (not shown;
qualitatively, the eigenmode is similar to those
shown in Fig.~\ref{fig_bdgeigenmode3} but without the third nodal line).
For large $D$ [$D\approx 36$, see Fig.~\ref{fig_bdgfrequency}(c)], 
the instability thus acquires some similarities
with a
radial roton instability (see below).

\begin{figure}
\vspace*{.2cm}
\includegraphics[angle=0,width=70mm]{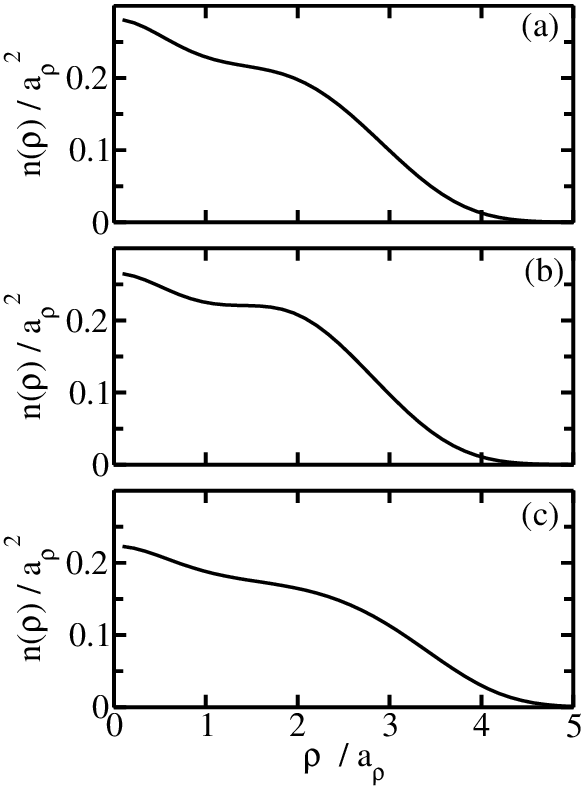}
\caption{
Integrated ground state density profiles $n(\rho)$ 
with ripple-like oscillations 
as a function of $\rho$ for $\lambda=9$, $b=0.2a_z$ and 
different $(D,A)$ combinations:
(a) $(D,A)=(50,0)$,
(b) $(D,A)=(68,8E_z)$ and
(c) $(D,A)=(99,12E_z)$.
The length scale 
at which the ripple-like oscillations occur is
to a good approximation
independent of the barrier height $A$ 
(see text for details). 
All three density profiles correspond to dynamically stable dipolar Bose gases
near the instability line [see Fig.~\protect\ref{fig_phasediagram}(c)].
The eigenmodes $\bar{f}$ 
of the corresponding
energetically lowest-lying $k=0$ Bogoliubov--de Gennes frequencies
are shown in Fig.~\protect\ref{fig_bdgeigenmode3}.
}\label{fig_ripple2}
\end{figure}

\begin{figure}
\vspace*{.2cm}
\includegraphics[angle=0,width=70mm]{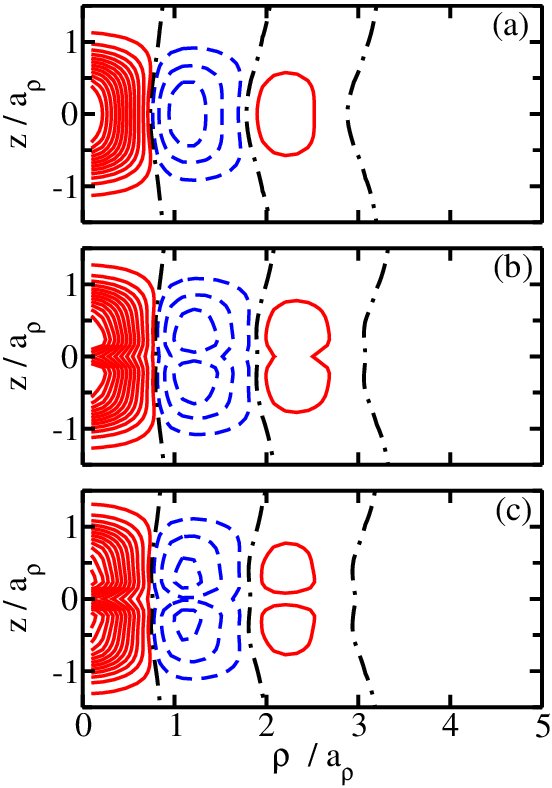}
\caption{
(Color online)
Examplary Bogoliubov--de Gennes eigenmodes $\bar{f}(\rho,z)$
for $k=0$, $\lambda=9$, $b=0.2a_z$
and different $(D,A)$ combinations:
(a) $(D,A)=(50,0)$,
(b) $(D,A)=(68,8E_z)$ and
(c) $(D,A)=(99,12E_z)$.
The contours are chosen equidistant, with solid and dashed lines
corresponding to positive and negative values 
of $\bar{f}$. The dash-dotted lines indicate the nodal lines of $\bar{f}$.
The corresponding integrated ground state densities $n(\rho)$
are shown in Fig.~\protect\ref{fig_ripple2}.}\label{fig_bdgeigenmode3}
\end{figure}

Lastly, we show that cylindrically symmetric systems
with modest $\lambda$ support,
in addition to density profiles of simple Gaussian shape and 
of red blood cell type shape,
density profiles with ripple-like oscillations.
Figures~\ref{fig_ripple2}(a)-(c) show examplary integrated density profiles
$n(\rho)$
with ripple-like oscillations 
for $\lambda=9$ and three different $(D,A)$ parameter combinations.
In general, we find density profiles with
ripple-like oscillations 
in a relatively small region near the dynamical instability points
for $\lambda=9$ but not for
$\lambda=8$ and $6$ 
[the parameter combinations that support ripple-like density profiles
fall into the S$_0$ region of 
the phase diagram shown in Fig.~\ref{fig_phasediagram}(c);
they are not indicated explicitly in Fig.~\ref{fig_phasediagram}(c)].
Ripple-like structures have very recently been predicted to exist
in 
dipolar systems with non-vanishing 
$s$-wave scattering length consisting of pancake-shaped stacks, where
each pancake is characterized by a large
aspect ratio $\lambda$, e.g., $\lambda=340$~\cite{kobe09}.
Ripple-like oscillations have also been predicted to exist for
dipolar systems with vortices~\cite{wils08}.
Here, we find ripple-like oscillations in the dynamically stable 
region of the phase diagram of pure
dipolar gases confined by
a cylindrically-symmetric trapping geometry with and without Gaussian barrier
and with modest aspect ratios $\lambda$.

The integrated density profiles shown in Fig.~\ref{fig_ripple2}(a)-(c)
for $A=0$ to $12E_z$
possess density oscillations with characteristic length 
scale of the order of $a_{\rho}$.
Figures~\ref{fig_bdgeigenmode3}(a)-(c) show the corresponding
eigenmodes $\bar{f}$ 
of the energetically lowest-lying $k=0$ 
Bogoliubov--de Gennes eigenfrequency.
The Gaussian barrier modifies the amplitude of the 
eigenmode near $z=0$ but otherwise leaves the overall structure of the
eigenmodes
unaffected.
All three eigenmodes possess roughly equally spaced nodal lines that
are to a good approximation independent of $z$, and
which can, roughly speaking, 
be characterized---just as the ripple-like oscillations---by 
$\rho_{\mathrm{node}} \approx n a_{\rho}$, where $n=1$, $2$ and $3$.
Using $\lambda=9$, 
it can be seen that $\rho_{\mathrm{node}}$
is determined by
$\rho_{\mathrm{node}} \approx \pi a_{\rho}/ \sqrt{\lambda}$
or $\rho_{\mathrm{node}}\approx  \pi a_z$~\cite{sant03,rone07}.  
The latter expressions reflect that
the condensate develops three-dimensional character, i.e., that it
``gets chopped up''
into smaller pieces along the $\rho$-direction. 
The dynamical instability that arises when 
$D$ is increased somewhat
compared to the $D$ values chosen in Figs.~\ref{fig_ripple2} and
\ref{fig_bdgeigenmode3}
[see also Fig.~\ref{fig_phasediagram}(c)] is thus identified as
a radial roton-like instability~\cite{sant03,rone07}.
Intuitively, one might expect that the 
length scale that characterizes the
ripple-like oscillations and the nodal lines of the Bogoliubov--de Gennes
eigenmodes would depend on the 
barrier height $A$ or, equivalently, the effective
aspect ratio $\lambda_{\mathrm{eff}}$. Figures~\ref{fig_ripple2} 
and \ref{fig_bdgeigenmode3}, however, suggest that the
system behavior is determined by $\lambda$ instead. 
We note, though, that the change of $\lambda_{\mathrm{eff}}$ 
as a function of $A$
is fairly small, implying that further studies are needed to 
determine the relevant length scale unambiguously.

\section{Summary}
\label{sec_conclusion}
We have investigated oblate dipolar Bose gases 
with vanishing $s$-wave scattering length
in a double well potential
as a function of the barrier height
$A$,
the mean-field strength $D$ and the aspect ratio $\lambda$.
Our stationary and dynamical 
mean-field calculations add two new aspects
to the already extensive list of intriguing behaviors
of dipolar Bose gases:
(i) We find an instability island immersed in a mechanically and dynamically
stable region of the phase diagram. This instability island can,
as has been outlined in Secs.~\ref{introduction} and \ref{sec_results},
be probed with present-day technology
and arises due to the going soft of a radial breathing mode-like frequency. 
(ii) We find structured ground state densities 
with ripple-like oscillations in the dynamically stable
region of the phase diagram for 
moderate aspect ratios and cylindrically-symmetric
confining geometries. These ripple-like oscillations exist,
for appropriately chosen parameter combinations, for vanishing
and non-vanishing Gaussian barriers; an increase of the mean-field strength
eventually
induces a radial roton-like instability.
In the future,
it will be interesting to investigate if the ground state densities
with ripple-like oscillations can be probed experimentally
using time-of-flight expansion techniques. 
Theoretically, this question can be addressed by
determining
the real time dynamics after turning off the confining potential
and by comparing the results with those for simple Gaussian
density profiles.
We emphasize that the emergence of the instability island and of 
density profiles with ripple-like oscillations is unique to
dipolar gases---these features are 
a direct signature of the long-range,
anisotropic dipole-diple interaction
and absent for $s$-wave interacting 
Bose gases.

Current experiments are directed at loading molecular
gases with appreciable electric dipole moment~\cite{ni08,deig08} 
into one-dimensional
lattices. The premise of these experiments is to exploit
the effectively two-dimensional geometry of each lattice site
as a stabilization mechanism.
In this context, several theoretical studies~\cite{wang08,klaw09,kobe09}
have investigated stacks of effectively two-dimensional
dipolar Bose gases. Common to these studies is the assumption
of a very tight confinement along the $z$-axis, i.e., a large
aspect ratio for each individual lattice site, and a comparatively
large spacing between individual lattice sites.
In those studies, roton and 
phonon instabilities emerge.
The studies presented here can be viewed as an alternative
approach to tackling the multi-site optical lattice system.
Using a single mean-field equation and 
working in a regime where
the
extent of the
cloud in one well is approximately
equal to the separation between clouds, the full system
dynamics is investigated for moderate aspect ratios.
Future studies will investigate the system behavior
as a function of the separation between sites, using 
a single mean-field equation and relatively small barrier
heights as well as a set of coupled mean-field equations;
these future studies 
will facilitate direct comparisons 
between the approach pursued here and 
Refs.~\cite{wang08,klaw09,kobe09}.

Support by the NSF through
grant PHY-0855332
is gratefully acknowledged.


\end{document}